\documentclass[12pt,preprint]{aastex}

\newcommand{\Ha}{H$\alpha$}
\newcommand{\Brg}{Br$\gamma$}
\newcommand{\dkl}{$\Delta (K - L)$}

\received{2002 September 4}
\accepted{2002 October 21}

\usepackage{emulateapj5}
\usepackage{apjfonts}
\journalinfo{Accepted by The Astrophysical Journal}
\submitted{Received 2002 September 4; accepted 2002 October 21}

\begin{document}
\title{Protoplanetary Disk Mass Distribution in Young Binaries}

\author{Eric L. N. Jensen}
\affil{Swarthmore College
Department of Physics and Astronomy, 
Swarthmore, PA 19081 USA
\email{ejensen1@swarthmore.edu}}
\author{Rachel L. Akeson}
\affil{Interferometry Science Center, Caltech,
MS 100-22,
1201 E. California Blvd.,
Pasadena, CA 91125 USA
\email{rla@ipac.caltech.edu}}

\shorttitle{Disk Mass Distribution in Young Binaries}
\shortauthors{Jensen and Akeson}

\begin{abstract}
  
  We present millimeter-wave continuum images of four wide
  (separations 210--800 AU) young stellar binary systems in the
  Taurus-Auriga star-forming region.  For all four sources, the
  resolution of our observations is sufficient to determine the mm
  emission from each of the components.  In all four systems, the
  primary star's disk has stronger millimeter emission than the
  secondary and in three of the four, the secondary is undetected;
  this is consistent with predictions of recent models of binary
  formation by fragmentation. The primaries' circumstellar disk masses
  inferred from these observations are comparable to those found for
  young single stars, confirming that the presence of a wide binary
  companion does not prevent the formation of a protoplanetary disk.
  Some of the secondaries show signatures of accretion (H$\alpha$
  emission and $K - L$ excesses), yet their mm fluxes suggest that
  very little disk mass is present.

\end{abstract}


\section{Introduction}

The majority of stars are in binary or multiple systems during both
the pre--main-sequence and main-sequence phases of stellar evolution
\citep[see, e.g., reviews by][]{Mathieu94,MathieuPPIV}. Therefore,
understanding the causes and effects of multiplicity is an essential
ingredient of complete models of both star and planet formation.
Circumstellar disks play a crucial role in both processes by providing
conduits for material to accrete onto the stars, and by providing
sites for planet formation.

Only in the last few years have observations of young binaries and
theories of binary formation both advanced to the point where direct
comparison of observations and theory is possible.  While such
comparisons still cannot definitively establish the mechanism(s) of
binary formation \citep[e.g.,][]{Clarke01}, increasingly there are
opportunities to test predictions of binary formation models with
observational data.  Current observations are consistent with the
formation of most low-mass binary systems by scale-free fragmentation
of a molecular cloud \citep{Clarke01,Ghez01}.  One notable prediction
of such models is that the more massive star in the binary should
always harbor the more massive circumstellar disk \citep{Bate00}.

We set out to test this prediction with the observations presented
here.  Previous work has shown that primary stars in young binary
systems have more active (and thus perhaps more massive) disks than
secondaries do.  Primaries tend to have stronger signatures of the
presence of circumstellar disks (red $K-L$ colors) and of accretion
from such disks (strong \Ha\ and \Brg\ emission) than secondaries do
\citep{PS97, Duc99, WG01, PM01}.  However, such observations reflect
the disk conditions near the stellar surfaces, where accretion takes
place and where the vast majority of the near-infrared flux is
produced.  These observations suggest a dependence of disk properties
on stellar mass in binaries, but in general these diagnostics
(especially \Ha\ equivalent width) are at most only marginally
correlated with disk mass \citep{bscg,OB95}.  Thus,
the optical and near-IR observations do not directly address the
question of total disk {\em masses\/} in these systems, making
comparison with theoretical predictions problematic.  A measurement of
the disk masses with millimeter interferometry is necessary for direct
comparison with theoretical predictions.  The work we present here has
the advantages that we (a) observed circumstellar disks at a
wavelength ($\lambda = 1.3$ mm) where they are likely to be optically
thin, to probe their total masses; (b) resolved both components in the
binaries; and (c) observed systems with mm fluxes that are typical of
average T Tauri stars, not anomalously large.

A few binaries have been previously observed with millimeter
interferometers, but the interpretation of their disk masses and
morphologies is complicated by the presence of unresolved additional
pairs in triple or quadruple systems (e.g., T Tau, \citealt{AKJ98}; UZ
Tau, \citealt{JKM96}).  Also, some of these observations have resolved
the circumbinary disks, but not the binary itself (GG Tau,
\citealt{KSB93}, \citealt{GDS99}; UY Aur, \citealt{Duvert98}).  Thus,
these observations do not reveal the distribution of disk mass between
the two stars in each system.  The observations we report here avoid
the complication of unresolved additional components (as far as we
know, except in the case of UX Tau B) and thus allow us to determine
the disk mass distribution in the circumstellar disks of four young
binary systems.

We present our observations in Section \ref{sec:observations} below.
We then discuss the disk properties implied by our observations and
other data (Sec.\ \ref{sec:modeling}) and the implications of these
results for our understanding of binary star formation and the
prospects for planet formation in binary systems (Sec.\ 
\ref{sec:discussion}).

\section{Observations}
\label{sec:observations}

We selected our targets from among known pre--main-sequence binaries
in Taurus-Auriga based on the following criteria: separation greater
than 1\farcs1, wide enough to resolve with the Owens Valley Millimeter
Array (which has a resolution of $1\farcs1 \times 0\farcs8$ for a full
track in the high-resolution configuration); and $\lambda=1.3$ mm
detection in the surveys of \citet{bscg} or \citet{OB95}.  We did not
restrict our sample to the brightest 1.3 mm sources, as limited
sensitivity has necessitated in the past.  We avoided triple and
quadruple systems in which the closer pair(s) would be unresolved in
our OVRO observations, since the presence of a companion with
separation less than 100 AU is well-correlated with decreased
millimeter emission \citep{bscg, OB95, JMF96}. We included UX Tau in
our sample since the A, B, and C components can be resolved by OVRO;
we did not realize until later that UX Tau B is itself a close binary.
Our sample consists of three binary systems (DK Tau, HK Tau, and V710
Tau) and one quadruple system (UX Tau); source properties are
summarized in Table \ref{table:sources}.  We note that the single-dish
mm fluxes of these systems (35--73 mJy) are fairly typical of T Tauri
star fluxes; the median flux among single stars in the samples of
\citet{OB95} and \citet{bscg} (including non-detections) is 36 mJy,
while the median flux among binary stars is somewhat lower.

We observed the four systems in Table \ref{table:sources} with the
Owens Valley millimeter-wave array on 11 and 19 January and 10
February 2002.  We obtained data in both the low-resolution array
(baselines 36--115 m; hour angle range $\pm$3 hours) and in the
high-resolution array (baselines 35--240 m; hour angle range 0--3
hours).  The correlator was configured with two wide-band channels
centered around 230 GHz, and narrow band channels centered around
$^{13}$CO(2$\rightarrow$1).  The combination of both sidebands resulted in a
total continuum bandwidth of 4 GHz.

During each observation, the array pointing was cycled among the four
sources to provide similar hour angle coverage for each.  The quasar
J0449+135 was used as the gain calibrator and 3c454.3 and 3c273 were
observed for use as flux calibrators.  As our observations contained
no measurements of primary flux calibrators (e.g., planets), we
estimate the absolute flux calibration uncertainty to be 20\%;
relative fluxes are unaffected by this uncertainty.  This uncertainty
is based on the scatter in other measurements of 3c454.3 and 3c273 at
OVRO that were calibrated using primary flux calibrators.  The gain
and flux calibrations were applied in the OVRO mma package.  Maps were
constructed and CLEANed in the MIRIAD package using a robust weighting
which resulted in a 1\farcs5 by 1\farcs3 beam.

\begin{figure}
  \plotone{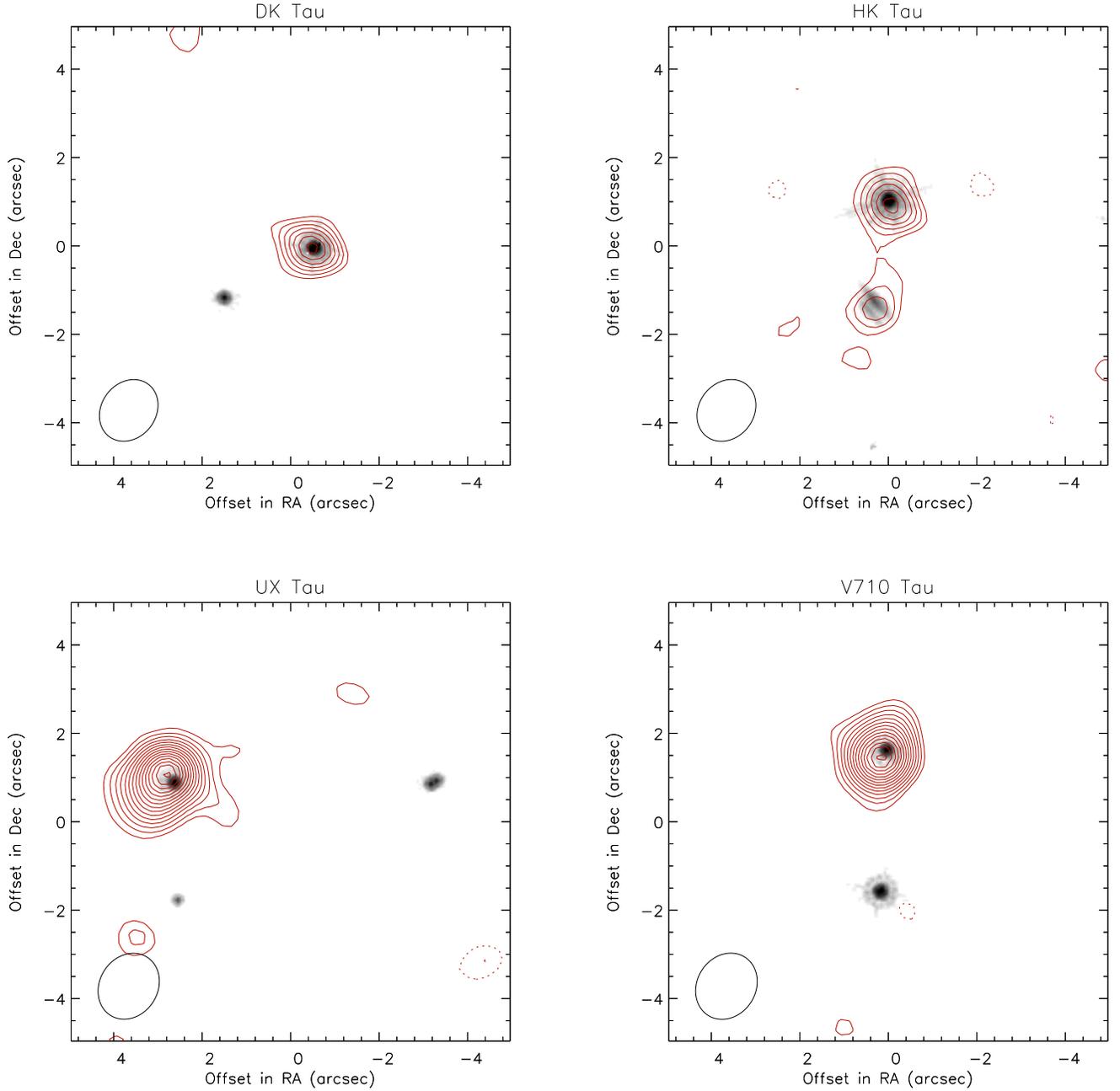}
  \caption{Millimeter emission from four young multiple systems.  The
    contours show the $\lambda = 1.3$ mm flux observed with the
    Owens Valley Millimeter Array overlaid on grayscale optical images
    from archival Hubble Space Telescope data.  In all cases the
    primary star dominates the disk emission from the system; only in
    HK Tau, the system with the mass ratio closet to one (see Table
    1), is there detectable emission from the secondary.  The contours
    are in steps of the RMS noise in the millimeter maps, 
    2.1--2.2 mJy, starting at the 3$\sigma$ contour; negative
    contours (dashed lines) have the same steps, starting at $-3\sigma$.  The beam size
    (1\farcs5 x 1\farcs3) is shown at lower left.
    \label{figure:fourpanel}}
\end{figure}

All four targets show at least one clearly-detected mm continuum
source (Table \ref{table:fluxes}; Figure \ref{figure:fourpanel}).
However, CO emission was not detected toward any of these sources;
3$\sigma$ limits are 0.7--0.8 Jy, $T_{\rm antenna} = 2.2$--2.7 K at a
resolution of 0.33 km/s; or 0.3--0.4 Jy, $T_{\rm antenna} = 1.0$--1.3
K when the spectra are binned to a resolution of 2 km/s. 
As discussed at the end of this section, our observations are less 
sensitive to emission on scales $> 5$\arcsec.

Three of the four systems show only one detected mm source.  To
determine which star is the source of the mm continuum emission, we
overlayed the mm maps onto archival Hubble Space Telescope (HST)
optical images of our sources taken in the F814W filter (similar to
Cousins $I$).  The absolute pointing accuracy of our observations is
0\farcs1 as measured by quasar observations, but uncertainties in
positions of the guide stars typically limit the accuracy of the
HST/WFPC2 coordinates to 0\farcs7 RMS \citep{Baggett02}.  Thus, to set
the absolute coordinates in the HST images more accurately we used
coordinates for our targets from the Hipparcos catalog (for UX Tau
only) or from the USNO A2.0 catalog \citep{Monet98}.  We used proper
motions from Hipparcos or \citet{JH79} to transform these coordinates to
the epoch of the OVRO observations.  Since the USNO A2.0 coordinates
do not resolve the binary systems, we assigned these coordinates to
the photocenter of the emission measured in the HST images.  The
results are shown in Figure \ref{figure:fourpanel}.  In all cases, the
source of the mm emission is unambiguous, since it aligns with the
optical emission of a star to within 0\farcs1--0\farcs2.  In each
case, the stronger mm emission comes from the primary star, and in
three of the four systems the secondary is undetected.

All of the mm detections except for UX Tau A are unresolved point
sources.  UX Tau A is marginally resolved; an elliptical Gaussian fit
to the millimeter emission gives a FWHM size of $1\farcs1 \times
0\farcs6$.  As this is at the limit of our resolution, the exact size
is uncertain; however, the clear difference between the source's
integrated flux and peak flux indicates that it is definitely
resolved.  The components of HK Tau A and B are consistent with
unresolved point sources, but there is some additional diffuse
emission between the two.

As can be seen in Table \ref{table:fluxes}, our fluxes appear to
be somewhat lower than the single-dish 1.3 mm fluxes for these
sources.  There are two possible explanations for this.  One is that
there is a systematic offset in the absolute flux calibration of the
two datasets.  As noted above, our absolute calibration is uncertain
by 20\%; \citet{bscg} and \citet{OB95} use different calibration
sources than we do and also cite a 20\% flux calibration uncertainty.
If we increase our fluxes by 20\% and take into account the random
uncertainties of each observation, then all of our measurements are
still less than the single-dish fluxes, differing by 1.8, 0.9, 0.06,
and 2.7 $\sigma$ for DK Tau, HK Tau, UX Tau, and V710 Tau
respectively.  If we shift our data up by 20\% and the single-dish
fluxes down by 20\%, then the differences are 0.8, $-$0.4, $-$1.1, and
1.6$\sigma$, with half of our measurements now being slightly higher
than the single-dish fluxes.

Alternatively, if the flux calibrations are correct, then the
discrepancy could be explained by the fact that the interferometric
measurements are not sensitive to extended emission.  Although a
discrepancy between interferometric and single-dish flux measurements
is expected for younger sources from which there is substantial envelope
emission, T Tauri sources are expected to have little if any envelope
component (see, e.g., HL Tau vs.\ L1551 IRS 5 in \citealt{Lay94}).
The percentage of the single-dish flux detected for our sources ranges
from 46 to 83\%.  Given the $uv$ coverage of our observations,
emission must have a size scale greater than 5\arcsec\ before half of
the emission is filtered out in the interferometric data.

\citet{PS97} have suggested that an extended circumbinary envelope
could replenish the disks in T Tauri binary systems, though it is
unclear how these envelopes could persist for so long
\citep[e.g.,][]{Clarke01}.  While our observations could be taken as
support for existence of such an extended component, we caution that
further observations are necessary to resolve the flux calibration
issue.

\section{Disk properties}
\label{sec:modeling}

In this section, we attempt to determine the properties of the
circumprimary and circumsecondary disks in our target systems.  We
first discuss what can be determined about the disks from existing
optical and infrared data, and then we address the disk properties as
revealed by our new observations.

\subsection{Disk properties inferred from optical and infrared data}

Resolved spectra of the individual stars in our target binaries are
presented in \citet{CK79, Magazzu91, HSS94, Monin98}, and
\citet{Duc99}.  We have quoted the range of literature spectral types
in Table \ref{table:sources}.  We adopted the spectral types given by
\citet{Duc99} for all systems except V710 Tau, for which we used the
spectral types from \citet{HSS94}.  In most cases, the different
references agree fairly well about the spectral types; however, two
cases merit some discussion.  For DK Tau, \citet{HSS94} list spectral
types of K7 for both components, raising the question of whether it is
clear which star is the primary. While \citet{HSS94} do not show their
spectra of DK Tau, the spectra shown in \citet{Monin98} clearly show
DK Tau B (their label) to be later in spectral type than DK Tau
A\null. Second, optical and infrared photometry in the literature
consistently show DK Tau A to be the brighter star.  Thus, the choice
of primary in this system seems clear.

In HK Tau, on the other hand, the spectra shown by \citet{Monin98}
appear to be quite similar in spectral type.  We adopt the M2 and M3
classifications of \citet{Duc99}, which are based on the
\citet{Monin98} spectra, but we note that the two stars are extremely
close in spectral type.  HK Tau B is much fainter, but most (if not
all) of this difference is due to its edge-on disk blocking the direct
starlight \citep{Stap98,Koresko98}.

The papers cited in the previous two paragraphs also present some
resolved photometry, as do \citet{WG01} ($K$ and $L$ for all sources),
\cite{MZ91} ($JHK$ for UX Tau), \citet{Koresko98} ($JHK$ for HK Tau),
and \citet{Stap98} ($JHK$, F606W, and F814W [HST filters similar to
$V$ and $I$] for HK Tau).

All of these systems were detected by IRAS and thus have substantial
mid- and far-infrared emission.  However, the IRAS observations do not
resolve the binaries and thus it is unclear which of the binary
components is the source of the infrared excess.  Indeed, prior to the
observations reported here, none of these systems had been resolved at a
wavelength longer than 3.6 \micron\ ($L$ band).

With the available {\em resolved\/} optical and infrared data, then,
the best tracers of the presence of circumstellar material are \Ha\ 
emission (presumed to arise from accretion onto the stellar surfaces)
and $K - L$ color (which can reveal the presence of an infrared excess
indicative of a disk).  These quantities are given in Table
\ref{table:sources}.

Among optical and near-infrared colors, $K-L$ is well-suited for
tracing circumstellar material because it is relatively unaffected by
interstellar reddening, and because the photospheric colors of
late-type stars have a relatively small range of values, minimizing
the effect of uncertainties in spectral typing on the calculation of a
color excess \citep{WG01}.  To determine the $K-L$ color excesses
$\Delta (K - L)$ given in Table \ref{table:sources}, we used the
photospheric colors of \citet{BB88}, which range from 0.10--0.20 for
the range of spectral types in our sample.

Both the \Ha\ emission and \dkl\ in Table \ref{table:sources} tell the
same story.  Stars classified as classical T Tauri Stars (CTTS) based
on \Ha\ equivalent width ($\ge 5$ \AA\ for K stars, $\ge 10$ \AA\ for
early M stars; see \citealt{Martin97}) also have significant
($>2.5$--3$\sigma$) \dkl\ color excesses.  The borderline CTTS V710
Tau B has only a borderline \dkl\ excess as well.  All four systems
thus have primary stars with evidence for disks. DK Tau and HK Tau
also have secondaries with evidence for disks, while V710 B is
marginal, and UX Tau B and C show little evidence of disk material.

\subsection{Disk properties inferred from millimeter observations}

As noted above, the most striking thing about the observed millimeter
fluxes from these systems is the how completely the primary stars
dominate the systems' mm emission.

\subsubsection{Comparison with optical and infrared disk properties}

\begin{figure}
  \plottwo{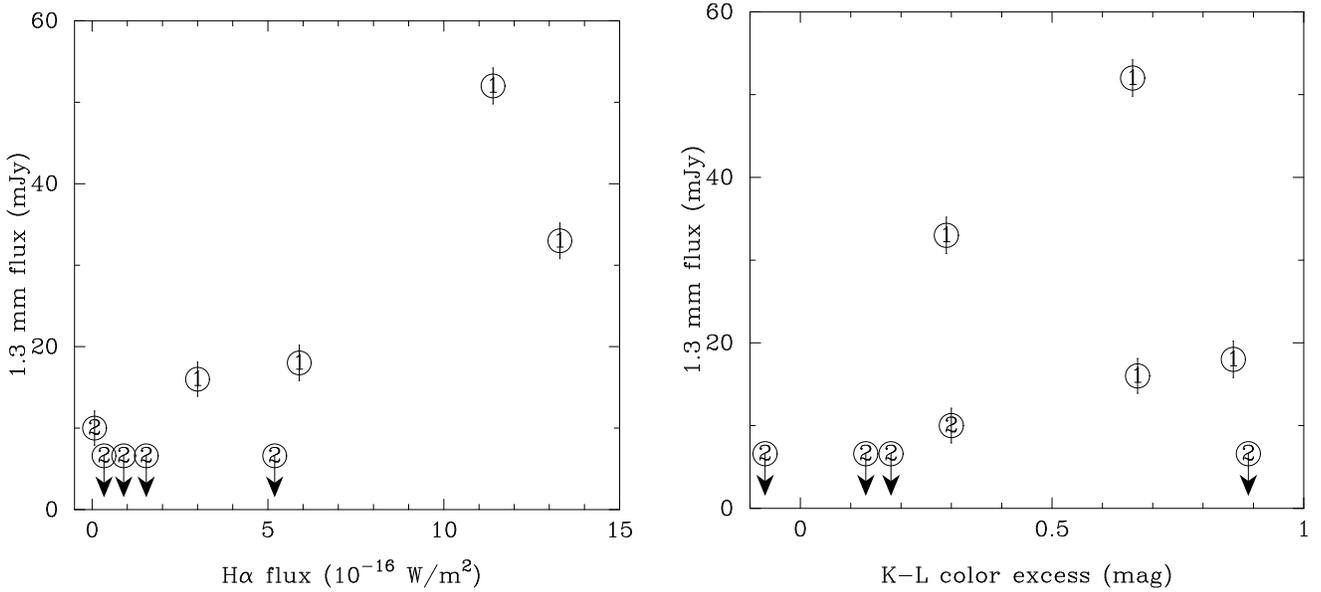}{f2b.eps}
  \caption{Left panel: Millimeter flux vs.\ \Ha\ emission-line
    flux.  Symbols with ``1'' indicate primary stars, and
    those with ``2'' denote secondary or tertiary stars.  There is no
    one-to-one correspondence between \Ha\ and mm flux, indicating
    that disk accretion (traced by \Ha) is not strongly correlated
    with disk mass.  Right panel: Millimeter flux vs.\ $K-L$ excess.
    Again, there is little correlation between $K-L$ excess, a tracer
    of inner disks, and the overall mass of the disk. \Ha\ fluxes are
    derived from the equivalent widths in Table \ref{table:sources}
    and  continuum fluxes estimated from spectra in \citet{Monin98} or
    $R$-band photometry (for V710 Tau only) in \citet{HSS94}.
    \label{figure:accretion}
    }
\end{figure}

There is not a one-to-one correlation between the properties discussed
in the preceding section and the millimeter fluxes in our observations
(Figure \ref{figure:accretion}).  While it appears that strong \Ha\ 
emission and $K-L$ excess are necessary conditions for the presence of
a millimeter detection, they are clearly not sufficient, since DK Tau
B shows strong \dkl\ and \Ha, but no millimeter emission.  This is
most likely due to the fact that the optical and near-infrared
emission arise in a relatively small region of the disk close to the
star, while the mm flux is more broadly distributed.  Thus, each
diagnostic has its advantages.  \Ha\ emission and $K-L$ excess can be
sensitive to relatively small disks that are undetected at millimeter
wavelengths, while mm emission is a better tracer of global disk
properties, especially disk mass.  Disk mass is the property that is
crucial for testing binary formation models, and it is there that we
now turn our attention.

\subsubsection{Disk mass estimates}
Millimeter-wavelength observations have often been used in conjunction
with optical and infrared data to model the spectral energy
distributions of T Tauri stars, with the optical and infrared data
providing some constraint on the disk temperature distribution, and
the millimeter data providing a tracer of optically thin material in
order to determine the disk mass \citep[e.g.,][]{bscg, OB95}.
However, there is not enough resolved mid- and far-infrared data
available for these binaries to justify detailed modeling of their
spectral energy distributions.
 
Limits on the disk masses can be estimated by assuming that all
emission is from optically-thin material with a given temperature.
These estimated disk masses (Table \ref{table:masses}) are a lower
limit to the true disk mass.  We use the canonical value for the dust
emissivity $\kappa_\nu$ from \citet{Hildebrand83} of 0.1 cm$^2$ g$^{-1}$
at $\lambda = 250$ \micron\ with a frequency scaling $\kappa_\nu
\propto \nu^{\beta}$ of $\beta = 1$.

The assumption that the disks are entirely optically thin at 1.3 mm is
unlikely to be correct for the primary stars with detected 1.3 mm
flux, and so these masses are lower limits to the true disk mass.
However, we note that these mass estimates are comparable to those
obtained from detailed models \citep{bscg, OB95} of stars with similar
mm fluxes, suggesting that the optically-thin approximation may be
reasonably good.  In their T~Tauri disk survey at $\lambda=2.7$~mm,
\citet{Dutrey96} estimated that optically thick emission accounted for
$\lesssim 10$\% of the total mass for 13 of 15 sources.  Using their
assumed disk model to extend these results to $\lambda=1.3$~mm,
optically thick emission at $\lambda=1.3$~mm accounts for less than
20\% of the total mass in 13 of 15 sources.

\subsubsection{Disk size estimates}

Alternatively, if we assume that the millimeter flux comes entirely
from optically {\em thick\/} material, we can estimate the minimum
size of the emitting region.  If a disk inclination value is assumed
(we used $\cos i = 0.5$) and the temperature is described as a power
law function of radius, $T(r)=T(r_0)*(r/r_0)^{-q}$, then for a given
flux, the outer radius for a completely optically thick disk depends
only on $T(r_0)$ and $q$.  For each source, we calculated the outer
radius using $T(r_0)=150$ K and $q=0.5$.  The calculated radii are
given in Table \ref{table:masses}.  Calculations using the $T(r_0)$
and $q$ values derived for each source in \cite{bscg} yielded similar
results.  These radii can be interpreted as lower limits to the true
disk radii, since it is likely that some of the emission is optically
thin.  The derived radii are similar to the limits found by
\citet{Dutrey96} for circumstellar disks around single T Tauri stars.

\section{Discussion}
\label{sec:discussion}

The 1.3 mm flux in these systems is clearly dominated by the primary
star.  The essential question for comparison of these observations
with predictions of theories of binary formation is whether or not
this larger flux indicates a larger disk mass around the primaries as
well. 

While some of the observed flux difference may be attributable to the
primaries' disks being hotter, we argue that the primaries must in
fact have more massive disks than the secondaries.  The primary and
secondary stellar effective temperatures differ by less than 10\% in
three of the four systems, and by 22--28\% in UX Tau. A temperature
difference of this magnitude is not sufficient to account for the
observed flux difference if the disks have similar masses and
opacities.

Another argument that could be made against the higher mm fluxes
resulting from higher-mass disks is that the disk inclinations of the
secondaries might be different from those of the primaries, with the
primaries presenting a larger projected surface area and thus a larger
mm flux from (perhaps) a similar disk mass.  The problem with this
hypothesis is that it requires a coincidence, namely that secondaries
happen to be more edge-on to Earth than the primaries in all four
cases.  Another problem arises in the case of HK Tau, where the
inclination of the secondary disk is known to be fairly edge-on to
Earth \citep[roughly 5\arcdeg;][]{Stap98} and the primary and
secondary disks are clearly not coplanar.  This is the system that has
perhaps the most edge-on secondary disk, yet it is the only system in
our sample in which the secondary was detected.  This argues strongly
against all the secondary disks being undetected due to an edge-on
geometry.  For equal-mass optically-thin primary and secondary disks,
only extreme inclination angles for the secondaries would result in
the primary/secondary flux ratios measured here.  Inclination angles
this extreme would probably obscure the optical emission from the
secondary, which is only the case for HK~Tau (Figure
\ref{figure:fourpanel}).  Finally, \citet{JDM00} find that among wide
binaries in general, primary and secondary disks tend to be aligned
with each other to within $\sim20$\arcdeg, making it unlikely that we
would see significant flux differences due to inclination differences
alone.

Thus, the secondaries not only have lower millimeter fluxes, but they
almost certainly have lower-mass disks.  This is in excellent
agreement with binary formation models of scale-free fragmentation
\citep[e.g.,][]{Bate00, Clarke01}.

\begin{figure}
  \plottwo{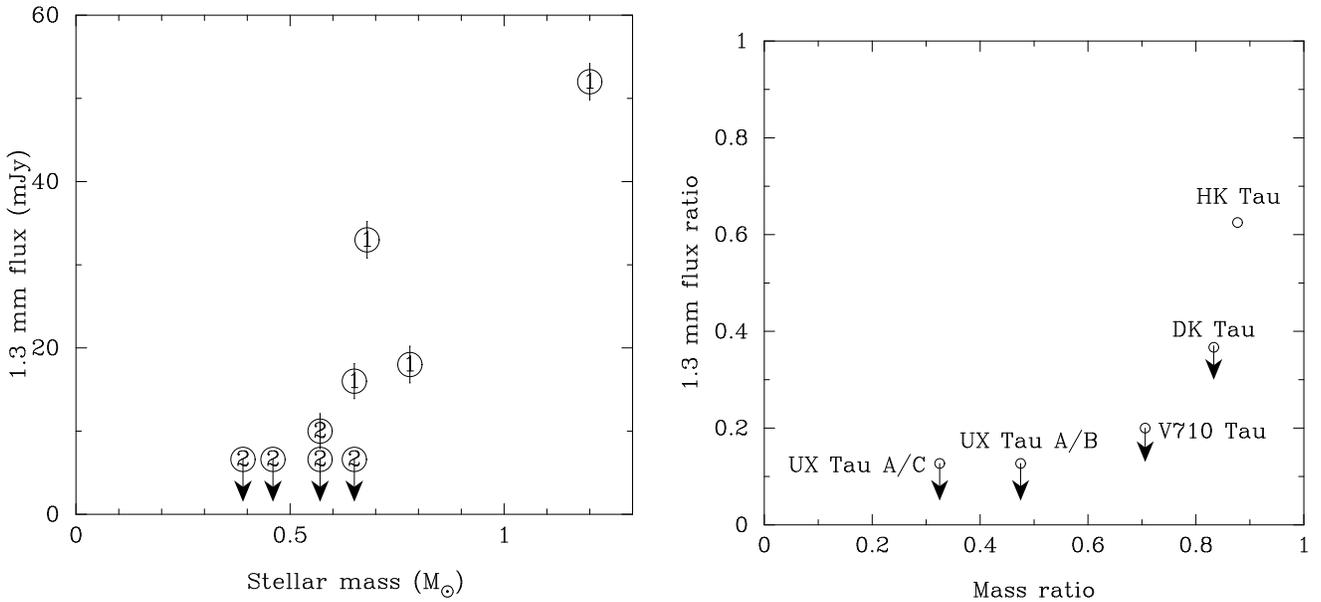}{f3b.eps}
  \caption{Left panel: Millimeter flux  vs.\
    stellar mass; symbols are as in Figure \ref{figure:accretion}.
    There is overlap between the primary and secondary mass
    distributions, with all primaries being detected while most
    secondaries, even those with masses comparable to some primaries,
     are undetected.  Right panel: The flux ratio
    from our millimeter observations vs.\ the stellar mass ratio.  Only when
    the stellar mass ratio is very close to one is the secondary
    detected.  Both panels suggest that the star's {\em relative\/}
    mass within the binary system is more important for its disk
    properties than its absolute mass.
    \label{figure:fluxvmass}
    }
\end{figure}

Examining the relationship between stellar mass and disk mass suggests
that the disk properties we observe are somehow related to the
dynamics of binary formation and evolution, and not just to the mass
of each individual star.  Some of the secondary stars in the systems
discussed here are roughly equal in mass to some of the primaries; for
example, HK Tau A and DK Tau B are both spectral type M1, and
therefore are very close in mass; V710 Tau A is also close at spectral
type M0.5.  Despite this similarity, HK Tau A and V710 Tau A, the more
massive stars in their respective systems, have detectable mm flux
(and therefore more massive disks) while DK Tau B has a very small mm
flux, below our sensitivity limit (Figure \ref{figure:fluxvmass}).
While our sample is small and thus we cannot draw definitive
conclusions about all young binaries, our data suggest that it is not
the individual stellar masses in a binary system that determine the
distribution of disk mass within the system.  Rather, the primary
(regardless of its absolute mass) retains a more substantial disk.  In
this scenario, the fact that both components in HK Tau have detectable
disks is consistent with the fact that it is the only system in which
the primary and secondary spectral types are virtually
indistinguishable.

It is also notable that the disk masses derived for the primaries are
similar to those derived for single T Tauri stars by \citet{bscg} and
\citet{OB95}, especially considering that our mass estimates are lower
limits.  This suggests that a wide ($> 200$ AU) binary companion does
not prevent the formation of a circumstellar disk that is near the
mass of the minimum mass solar nebula, consistent with the conclusions
of \citet{JMF96}.  In fact, this result strengthens the prospects for
planet formation in wide binaries, since it suggests that most of the
disk mass in these systems resides in a single, more massive disk
rather than two smaller disks, thus providing a larger reservoir of
material for planet formation around the primary.

The upper limits on the mm flux from the secondaries do not, in and of
themselves, set these stars apart from single stars; roughly half of
the single stars in the sample of \citet{OB95} were not detected at
1.3 mm.  What is striking, as emphasized before, is the comparison
between the mm emission of the primary and that of the secondary.
This is especially notable since a comparison of primary and secondary
disks within a given binary system is effectively controlled for the
effects of age and formation environment.  These factors allow for the
persistence of a substantial disk around the primaries in our sample,
and yet most of the secondaries are relatively devoid of disk material.

Several studies (see, e.g., the review by \citealt{PM01}) have shown
that T Tauri binaries tend to occur in matched pairs, i.e.\ it is much
more common to find classical T Tauri stars (CTTS) paired with other
CTTS and weak-lined T Tauri stars (WTTS) paired with other WTTS than
it is to find mixed CTTS/WTTS pairs.  The fact that both stars in a
system are CTTS does not mean that their disks are similar, however.
While \Ha\ traces disk accretion, it is not a good tracer of disk mass
\citep[e.g.,][]{bscg}.  Our results here reinforce the disparity
between accretion diagnostics and disk mass: while most of the
secondaries are CTTS, they have very low-mass disks (Table
\ref{table:masses}) and thus their disks are fairly different from
those around the primaries.  The most notable example is DK Tau B,
with an \Ha\ emission equivalent width of 118 \AA\ but no detectable
$\lambda= 1.3$ mm emission.  Thus, while both primary and secondary in
a binary system are often CTTS, primaries and secondaries can have
very different disk masses.

\section{Conclusions}

We have presented $\lambda=1.3$ mm continuum images of four young
binary systems, showing that the primary star has a more massive disk
in all cases; this is consistent with leading models of binary
formation by fragmentation.  These circumprimary disks are comparable
in mass to those found around single T Tauri stars, indicating that
the presence of a wide binary formation does not prevent the formation
and survival of a disk massive enough to form a solar system like our
own.  In some systems, the secondary star shows evidence of strong
accretion, but no detectable mm emission, indicating that the
reservoir of accreting material has a relatively low mass.  None of
the systems observed has detectable $^{13}$CO(2$\rightarrow$1)
emission at the sensitivity limit of our observations.

\acknowledgments

This work was performed in part at the Interferometry Science Center,
California Institute of Technology. The Owens Valley millimeter-wave
array is supported by NSF grant AST 96-13717.  ELNJ gratefully
acknowledges the support of the National Science Foundation's Life in
Extreme Environments program through grant AST 99-96278, and
Swarthmore College through a James Michener Fellowship.  We thank the
anonymous referee, whose detailed comments improved the presentation
of this work.  We thank Rabi Whitaker for a careful reading of the
paper.

\clearpage

\clearpage

\begin{deluxetable}{lccccccccc}
\tablewidth{0in}
\tablecaption{Young Binaries Observed\label{table:sources}}
\tablehead{
 \colhead{System} & &
    \colhead{Proj.} &
    \colhead{Pos.} & \colhead{Lit.} & \colhead{Adopted} &
    \colhead{H$\alpha$} & $K-L$  & $\Delta(K-L)$ & \colhead{Stellar}  \\
   & & \colhead{Sep.} &
    \colhead{Ang.} & \colhead{spectral} & \colhead{spectral} &
    \colhead{EW} & &&\colhead{mass} \\
    &&& & \colhead{types} &  \colhead{type} &\colhead{(\AA)} &&&\colhead{($M_\sun$)} 
}
\startdata
DK Tau    & A &      &     & K7--K9  & K9  &  \phn31\phd\phn  & $0.99 \pm 0.04$ &
\phs0.86 & 0.78 \\
        & B & 2\farcs8   & 115\arcdeg  & K7--M1  & M1  & 118\phd\phn  & $1.04
    \pm 0.08$ & \phs0.89 &0.65 \\[2mm]
  HK Tau    & A &      &       & M1  & M1  &  \phn50\phd\phn &
$0.82 \pm 0.04$ & \phs0.67 &0.65 \\
   & B & 2\farcs4   & 175\arcdeg  & M2  & M2  &  \phn12.5 & $0.46 \pm 0.12$&  \phs0.30
    & 0.57 \\[2mm]
UX Tau    & A &      &     & K2--K5  & K4  & \phn\phn9.5 & $0.76 \pm 0.06$
& \phs0.66&1.2  \\
 & B\tablenotemark{a} & 5\farcs9  & 269\arcdeg
& M1--M2  & M2  & \phn\phn4.5 & $0.09 \pm 0.06$&  $-0.07$ &0.57  \\
 & C & 2\farcs7   & 181\arcdeg  & M3--M5  & M3  & \phn\phn8.5 & $0.38 \pm 0.12$& \phs0.18 &0.35 \\[2mm]
V710 Tau  & A &      &     &  M0.5--M1 &  M0.5 & \phn89\phd\phn & $0.44 \pm 0.04$ & \phs0.29 &0.68 \\
 & B & 3\farcs2   & 177\arcdeg  &  M2--M3 &  M2.5 & \phn11\phd\phn & $0.31 \pm 0.04$
& \phs0.13 & 0.48 \\
\enddata

\tablecomments{Position angle and separation values are from
  \citet{Leinert93} and are relative to the primary star.  For V710
  Tau, component A here is the {\em optical\/} primary, though
  component B (V710 Tau S) is brighter at 2.2 \micron. Adopted spectral types
  and \Ha\ equivalent widths are from \citet{HSS94} for V710 Tau and
  \citet{Duc99} for all other stars; $K-L$ values are from
  \citet{WG01}.  Spectral types were used to derive effective
  temperatures based on Table 2 of \citet{Luhman99} for the M stars
  and based on the dwarf temperature scale in \citet{Allen91} for the
  K stars.  The masses were then estimated from the \citet{Baraffe98}
  models assuming an age of 3 Myr, since \citet{Luhman00} found a
  sample of young stars in Taurus to lie between the 1 and 3 Myr
  tracks of the \citet{Baraffe98} models.  Although this method does
  not produce precise values for the individual masses, the relative
  masses should be relatively well determined.}
\tablenotetext{a}{UX Tau B is a 0\farcs14 binary
  (Duch\^ene 1999), unresolved in our data but visible in
  the HST data in Figure \ref{figure:fourpanel}. }

\end{deluxetable}

\begin{deluxetable}{rcccc}
\tablewidth{0pt}
\tablecaption{Millimeter fluxes\label{table:fluxes}}
\tablehead{
\colhead{Source} & & \colhead{1.3 mm} & 
\colhead{Source} & \colhead{1.3 mm} \\ 
&& \colhead{OVRO flux} & \colhead{size} &
\colhead{single-dish}  \\
&& \colhead{(mJy)\tablenotemark{a}}  &  \colhead{(arcsec)\tablenotemark{b}}& \colhead{flux (mJy)\tablenotemark{c}}
}
\startdata
\\[-5mm]
DK Tau & A & $18 \pm 2.2$ & point & $35 \pm 7$ \\
 &B & $< 6.6$ & \nodata & \\[5mm]
HK Tau & A & $16 \pm 2.1$ & point  & $41 \pm 5$ \\
& B & $14 \pm 2.1$ & point \\[5mm]
UX Tau & A & $52 \pm 2.2$ & $1\farcs1 \times 0\farcs6$  &
$63 \pm 10$ \\
& B & $<6.6$  & \nodata \\
& C & $<6.6$  & \nodata \\[5mm]
V710 Tau & A & $33 \pm 2.2$ & point  & $60 \pm 7$ \\ 
& B & $<6.6$ & \nodata \\[2mm]
\enddata
\tablenotetext{a}{Errors given are the RMS noise in the maps,
  and do not include the absolute flux calibration
  uncertainty of 20\%.  Upper limits given are $3 \sigma$.}
\tablenotetext{b}{The size
  given for UX Tau A is for a Gaussian fit to the emission.  In the fit to HK~Tau,
  there is some diffuse residual emission between the two sources.}
\tablenotetext{c}{ The
  single-dish fluxes are taken from \citet{OB95} or
  \citet{bscg} and have a beam size of 11\arcsec, encompassing all the
  stars in each system. The uncertainties do not include the absolute
  flux calibration uncertainty. }
\end{deluxetable}

\begin{deluxetable}{lccc}
\tablecaption{Disk mass and radius lower limits\label{table:masses}}
  \tablewidth{0pt}
\tablehead{ & \multicolumn{2}{c}{Optically thin} & \colhead{Optically thick}\\
  \colhead{Source} & \multicolumn{2}{c}{disk mass ($10^{-3}$ M$_\sun$)}
  & \colhead{disk radius}\\
& \colhead{$T = 15$~K} & \colhead{$T = 30$~K}
& (AU)
}
\startdata
DK Tau A & 3.4  & 1.7  & \phn9.1 \\
\phantom{DK Tau }B & $<1.2 $ & $<0.6 $ & \nodata \\
HK Tau A & 3.0  & 1.5  & \phn8.4 \\
\phantom{HK Tau }B & 1.9  & 0.9  & \phn6.1 \\
UX Tau A & 9.7  & 4.9  & 18.5 \\
\phantom{UX Tau }B & $<1.2 $ & $<0.6 $ & \nodata \\
\phantom{UX Tau }C & $<1.2 $ & $<0.6 $ & \nodata\\
V710 Tau A & 6.2  & 3.1  & 13.6\\
\phantom{V710 Tau }B & $<1.2 $ & $<0.6 $ & \nodata\\
\enddata
\end{deluxetable}


\begin{thebibliography}{}

\bibitem[Akeson, Koerner, \& Jensen(1998)]{AKJ98} Akeson, 
R.~L., Koerner, D.~W., \& Jensen, E.~L.~N.\ 1998, \apj, 505, 358 

\bibitem[Allen(1991)]{Allen91} Allen, C.\ W., 1991, Astrophysical
  Quantities, 3rd edition (London: The Athlone Press)

\bibitem[Baggett et al.(2002)]{Baggett02} Baggett, S., et al.\ 2002, HST WFPC2
  Data Handbook, v.4.0, ed. B.~Mobasher (Baltimore:STScI), Sec.\ 5.4

\bibitem[Baraffe et al.(1998)]{Baraffe98}   Baraffe, I., Chabrier, G., Allard,
F., \& Hauschildt, P.~H.\ 1998, \aap, 337, 403
  
\bibitem[Bate(2000)]{Bate00} Bate, M.~R.\ 2000, \mnras, 314, 
33 

\bibitem[Beckwith et al.(1990)]{bscg} 
Beckwith, S.~V.~W., Sargent, A.~I., Chini, R.~S., \& G\"usten, R.\ 1990, 
\aj, 99, 924 (BSCG)

\bibitem[Bessell \& Brett(1988)]{BB88} Bessell, M.~S.~\& 
Brett, J.~M.\ 1988, \pasp, 100, 1134 

\bibitem[Clarke(2001)]{Clarke01} Clarke, C.~J.\ 2001, in Birth and
  Evolution of Binary Stars, Proceedings of IAU Symposium 200, 346

\bibitem[Cohen \& Kuhi(1979)]{CK79} Cohen, M.~\& Kuhi, 
L.~V.\ 1979, \apjs, 41, 743 

\bibitem[Duch{\^ e}ne et al.(1999)]{Duc99}
  Duch{\^ e}ne, G., Monin, J.-L.,
  Bouvier, J., \& M{\' e}nard, F.\ 1999, \aap, 351, 954 

\bibitem[Dutrey et al.(1996)]{Dutrey96} Dutrey, A., Guilloteau, 
S., Duvert, G., Prato, L., Simon, M., Schuster, K.\ \& Menard, F.\ 1996, 
\aap, 309, 493


\bibitem[Duvert et al.(1998)]{Duvert98} Duvert, G., Dutrey, A.,
  Guilloteau, S., Menard, F., Schuster, K., Prato, L., \& Simon, M.\ 
  1998, \aap, 332, 867
  
\bibitem[Ghez(2001)]{Ghez01} Ghez, A.~M.\ 2001, in Birth and Evolution
  of Binary Stars, Proceedings of IAU Symposium 200, 210
  
\bibitem[Guilloteau, Dutrey, \& Simon(1999)]{GDS99} Guilloteau, S.,
  Dutrey, A., \& Simon, M.\ 1999, \aap, 348, 570
  
\bibitem[Hartigan, Strom, \& Strom(1994)]{HSS94} Hartigan, 
P., Strom, K.~M., \& Strom, S.~E.\ 1994, \apj, 427, 961 

\bibitem[Hildebrand(1983)]{Hildebrand83} Hildebrand, R.~H.\ 1983, 
\qjras, 24, 267 

\bibitem[Jensen, Donar, \& Mathieu(2000)]{JDM00} Jensen, 
E.~L.~N., Donar, A.~X., \& Mathieu, R.~D.\ 2000, in Birth and
Evolution of Binary Stars, Poster Proceedings of IAU Symposium 200, 85P

\bibitem[Jensen, Mathieu, \& Fuller(1996)]{JMF96} Jensen, E. L. 
  N., Mathieu, R. D. \& Fuller, G. A. 1996, \apj, 458, 312 

\bibitem[Jensen, Koerner \& Mathieu(1996)]{JKM96} Jensen, E.\ 
  L.\ N., Koerner, D.\ W.\ \& Mathieu, R.\ D.\ 1996, \aj, 111, 2431 
  
\bibitem[Jones \& Herbig(1979)]{JH79} Jones, B.~F.~\& 
Herbig, G.~H.\ 1979, \aj, 84, 1872

\bibitem[Koerner, Sargent, \& Beckwith(1993)]{KSB93}  Koerner,
D.~W., Sargent, A.~I., \& Beckwith, S.~V.~W.\ 1993, \apjl, 408, L93

\bibitem[Koresko(1998)]{Koresko98} Koresko, C.~D.\ 1998, \apjl, 
507, L145 

\bibitem[Lay et al.(1994)]{Lay94} 
Lay, O.~P., Carlstrom, J.~E., Hills, R.~E., \& Phillips, T.~G.\ 1994, 
\apjl, 434, L75 

\bibitem[Leinert et al.(1993)]{Leinert93} Leinert, C., Zinnecker,
H., Weitzel, N., Christou, J., Ridgway, S.~T., Jameson, R., Haas, M., \&
Lenzen, R.\ 1993, \aap, 278, 129

\bibitem[Luhman(1999)]{Luhman99} Luhman, K.~L.\ 1999, \apj, 525,
466

\bibitem[Luhman(2000)]{Luhman00} Luhman, K.~L.\ 2000, \apj, 544,
1044

\bibitem[Magazzu, Mart\'\i n, \& Rebolo(1991)]{Magazzu91} Magazzu,
A., Mart\'\i n, E.~L., \& Rebolo, R.\ 1991, \aap, 249, 149

\bibitem[Mart\'\i n(1997)]{Martin97} Mart\'\i n, E.~L.\ 1997, \aap, 321, 
492 

\bibitem[Mathieu(1994)]{Mathieu94}  Mathieu, R.~D.\ 1994, \araa,
32, 465

\bibitem[Mathieu et al.(2000)]{MathieuPPIV} Mathieu, R.\ D., Ghez, A.\ 
  M., Jensen, E.\ L.\ N.\ {\&} Simon, M.\ 2000, Protostars and Planets
  IV (Tucson: University of Arizona Press; eds Mannings, V., Boss,
  A.P., Russell, S.\ S.), p.\ 703

\bibitem[Monet et al.(1998)]{Monet98} Monet, D., Bird, A., Canzian, B., Dahn, C., Guetter, H.,
  Harris, H., Henden, A., Levine, S., Luginbuhl, C., Monet, A. K. B.,
  Rhodes, A., Riepe, B., Sell, S., Stone, R., Vrba, F., \& Walker,
  R., The USNO-A2.0 Catalogue, (Washington DC: U.S. Naval Observatory)
  
\bibitem[Moneti \& Zinnecker(1991)]{MZ91} Moneti, A.~\& 
Zinnecker, H.\ 1991, \aap, 242, 428 

\bibitem[Monin, M{\' e}nard, \& Duch{\^ e}ne(1998)]{Monin98} Monin,
J.-L., M{\' e}nard, F., \& Duch{\^ e}ne, G.\ 1998, \aap, 339, 113

\bibitem[Osterloh \& Beckwith(1995)]{OB95} Osterloh, M.~\&
Beckwith, S.~V.~W.\ 1995, \apj, 439, 288

\bibitem[Prato \& Monin(2001)]{PM01} Prato, L.~\& Monin, J.\ 
2001, in Birth and Evolution of Binary Stars, Proceedings of IAU Symposium 200, 313 

\bibitem[Prato \& Simon(1997)]{PS97} Prato, L.~\&
  Simon, M.\ 1997, \apj, 474, 455

\bibitem[Stapelfeldt et al.(1998)]{Stap98} Stapelfeldt, K.~R., 
Krist, J.~E., Menard, F., Bouvier, J., Padgett, D.~L., \& Burrows, C.~J.\ 
1998, \apjl, 502, L65 

\bibitem[White \& Ghez(2001)]{WG01} White, R.~J.~\&
  Ghez, A.~M.\ 2001, \apj, 556, 265

\end{thebibliography}
\end{document}